\documentclass[superscriptaddress,showpacs,dvips]{revtex4}
\usepackage{amssymb}
\usepackage{epsfig}
\usepackage{graphicx}
\usepackage{subfigure}
\usepackage{hyperref}

\begin{document}

\def\ps{p\hspace{-0.13cm}\slash}

\title{Dynamical fermion mass generation and exciton spectra in graphene}
\author{Chun-Xu Zhang}
\email{cxzhang.zhang373@gmail.com, cxzhang@nudt.edu.cn} \affiliation{Department of
Physics, National University of Defense Technology, Changsha, Hunan 410073, P.
R. China} \affiliation{Interdisciplinary Center for Theoretical
Study, University of Science and Technology of China, Hefei, Anhui
230026, P. R. China}
\author{Guo-Zhu Liu}
\email{gzliu@ustc.edu.cn} \affiliation{Department of Modern Physics,
University of Science and Technology of China, Hefei, Anhui 230026,
P. R. China} \affiliation{Institut f$\ddot{u}$r Theoretische Physik,
Freie Universit$\ddot{a}$t Berlin, Arnimallee 14, D-14195 Berlin,
Germany}
\author{Ming-Qiu Huang}
\email{mqhuang@nudt.edu.cn} \affiliation{Department of Physics,
National University of Defense Technology, Changsha, Hunan 410073, P. R.
China}

\begin{abstract}
The Coulomb interaction between massless Dirac fermions may induce
dynamical chiral symmetry breaking by forming excitonic pairs in
clean graphene, leading to semimetal-insulator transition. If the
Dirac fermions have zero bare mass, an exact continuous chiral
symmetry is dynamically broken and thus there are massless Goldstone
excitons. If the Dirac fermions have a small bare mass, an
approximate continuous chiral symmetry is dynamically broken and the
resultant Goldstone type excitons become massive, which is analogous
to what happens in QCD. In this paper, after solving Dyson-Schwinger
gap equation in the presence of a small bare fermion mass, we found
a remarkable reduction of the critical Coulomb interaction strength
for excitonic pair formation and a strong enhancement of dynamical
fermion mass. We then calculate the masses of Goldstone type
excitons using the SVZ sum rule method and operator product
expansion technique developed in QCD and find that the exciton
masses are much larger than bare fermion mass but smaller than
the width of dynamical fermion mass gap. We also study the spin 
susceptibilities and estimate the masses of non-Goldstone type 
excitons using the same tools.
\end{abstract}


\maketitle

%%%%%%%%%%%%%%%%%%%%%%%%%%%%%Main Body%%%%%%%%%%%%%%%%%%%%%%%%%%%%%%%%%%%%%

\section{Introduction}\label{Sec1}

Graphene is a recently fabricated two-dimensional electron system.
It has attracted a great deal of research activities since it
exhibits many interesting properties and also has remarkable
potential technical applications \cite{CastroNeto, DasSarma}. Its
fundamental low-energy degrees of freedom are two-dimensional
massless Dirac fermions, which obey relativistic Dirac equation. Due
to the special linear dispersion of Dirac fermions, the density of
states vanishes linearly near the touching points of conduction and
valence bands. Therefore, graphene is classified as a semi-metal and
the Coulomb interaction between massless Dirac fermions is
unscreened.

The semi-metal ground state of graphene is stable when the Coulomb
interaction is weak. Its microscopic action respects an exact
continuous chiral symmetry. Due to the masslessness of Dirac
fermions, graphene exhibits highly unusual behaviors, such as
minimum conductivity \cite{CastroNeto, DasSarma}, quantum Hall
effect \cite{Gusynin2007}, and marginal Fermi liquid behavior
\cite{DasSarma2007}. However, when the Coulomb interaction is
sufficiently strong, a finite mass gap will be dynamically generated
due to excitonic pairing instability \cite{Khveshch2004,
Gorbar_Gusynin2002, Liu2009, Hands2008, Drut2009,Araki2010,Son2007, Gamayun2010}. As a
consequence, the continuous chiral symmetry is broken and the
graphene undergoes a quantum phase transition from semi-metal to
excitonic insulator. The mechanism underlying this phase transition
can be considered as a concrete realization of the non-perturbative
phenomenon of dynamical chiral symmetry breaking (DCSB) that was
originally proposed by Nambu and Jona-Lasinio in the context of
particle physics \cite{Nambu_Jona-Lasinio1961}.

Once a finite fermion mass is generated dynamically, both the ground
state and the low-energy elementary excitations of graphene are
fundamentally changed. The ground state becomes insulating, and the
only low-energy excitations are massless Goldstone bosons induced by
dynamical breaking of exact continuous chiral symmetry. These bosons
are composed of quasi-particles (fermion) and quasi-holes
(anti-fermion), called excitons, and dominate the low-energy
behaviors of graphene.

Note that the excitonic pairing triggered by Coulomb interaction is
not the only way to generate a fermion mass gap. For example, the
Kekule type distortion can open a small fermion gap \cite{Chamon,
Hou_Chamon2007, Dillenschneider2008}. Moreover, the spin-orbit
interaction may generate a small fermion gap that has opposite sign
at $K$ and $K'$ points \cite{Kane}, which was proposed to be able to
turn graphene into a topological insulator \cite{Kane}. Other gap
generating mechanisms (for instance, geometry confinement) are also
possible \cite{Castro-Neto2009}. From an experimental point of view,
a finite fermion mass gap is expected to be present if there is an
exponential suppression of certain observable quantities, such as
specific heat and susceptibility, at low temperature. However, the
graphene may display different behaviors if the fermion gap is
generated by different mechanisms. To see this clearly, it is
helpful to make a symmetry analysis since different gap generating
mechanisms are usually associated with different symmetry breaking
patterns.

When there is no excitonic pairing instability, the fermion mass may
be generated by other mechanisms such as Kekule distortion. In this
case, the continuous chiral symmetry is broken explicitly, rather
than dynamically, and there are no Goldstone type excitons. If the
fermion mass is completely generated by excitonic pairing
instability, then there are massless Goldstone type excitons. The
third possibility is that more than one mechanisms are important.
Before the Coulomb interaction is turned on, the Dirac fermions may
already have a finite mass due to certain mechanism. If this bare
mass is small, the system possesses an approximate continuous chiral
symmetry. When the Coulomb interaction is turned on, the Dirac
fermions can acquire further mass gap due to excitonic pair
formation. Once this happens, the approximate continuous chiral
symmetry is dynamically broken \cite{Weinberg}. As a consequence,
the Goldstone type excitons have finite masses \cite{Weinberg}.

We believe that the third possibility is of particular interests for
two reasons. First, this possibility can happen in realistic
graphene materials. For instance, when Coulomb interaction and
Kekule distortion (or other mechanisms that can induce bare fermion
mass) are both present, there may be dynamical breaking of
approximate chiral symmetry. Second, the physical picture of this
possibility is very similar to what happens in QCD. In QCD, it is
well-known that the $u$ and $d$ quarks have small bare masses. When
these quarks acquire dynamical mass due to the formation of chiral
condensate, the approximate chiral symmetry is dynamically broken
and there appear massive Goldstone bosons, which are identified as
the $\pi$ mesons \cite{Weinberg}. The massive Goldstone type
excitons generated in graphene correspond to the massive $\pi$
mesons in QCD.

In this paper, we study the dynamical breaking of approximate chiral
symmetry in graphene. We first assume a bare constant mass for the
Dirac fermions, and then calculate the dynamical fermion mass by
solving the corresponding Dyson-Schwinger (DS) gap equation. From
the solutions, the chiral condensate can be easily obtained. We will
show that a small bare fermion mass can greatly catalyze the
formation of excitonic pairs. The critical Coulomb interaction
strength for excitonic pairing instability is reduced to zero, while
the dynamical fermion gap is significantly enhanced.

The dynamical fermion gap breaks the approximate chiral symmetry, so
there appear massive Goldstone type excitons. Now a natural problem
is to estimate the masses of these excitons, which can help to
understand the low-energy excitations in the dynamical symmetry
breaking phase. The close analogy between the approximate DCSB
behaviors in graphene and in QCD allows us to calculate the exciton
mass using the tools developed in QCD. In particular, we will use
the Shifman-Vainshtein-Zakharov (SVZ) sum rule method and the
operator product expansion (OPE) technique. OPE was proposed
independently by Wilson \cite{Wilson1969} and Kadanoff
\cite{Kadanoff1969}. Wilson's motivation came from the urgency to
study the problem of strong interactions associated with the nuclear
force, while Kadanoff applied the OPE to understand critical
phenomena in condensed matter physics. Nowadays, OPE is known to be
a powerful tool of quantum field theory describing elementary
particle physics \cite{Shifman1998} and condensed matter physics
\cite{Tsvelik1998}. SVZ sum rule technique was developed with the
aim to understand strong interaction \cite{SVZ1979a,SVZ1979b}. Among
its wide applications in hadronic physics, SVZ sum rule is
particularly powerful in calculating the masses of mesons and
baryons that are bound states of quarks and/or anti-quarks. These
methods were also applied to condensed matter systems, including
degenerate electron gas \cite{Rosenfelder1985} and cold atom system
\cite{Braaten2008}. In this paper, we calculate the masses of
excitons using these analytical tools.

These Goldstone type excitons are associated with the onset of
charge density wave (CDW) or spin density wave (SDW) orders. In
addition to these excitations, there are also collective spin
triplet excitons, which are not related to any symmetry breaking.
The masses of these non-Goldstone type excitons will also be
calculated by means of the same SVZ sum rule and OPE methods. 
We hope these exciton modes
could be found in future experiments (such as inelastic neutron
scattering) in clean graphene sheet.

The rest of this paper is organized as follows. In Sec.\ref{Sec2},
we give the continuum model of two-dimensional Dirac fermions
interacting through Coulomb potential. We study the DS gap equation
in the presence of small bare fermion mass and calculate the chiral
condensate using the dynamical fermion mass. In Sec.\ref{Sec3}, we
study the Goldstone type excitons and calculate their masses using
SVZ sum rule and OPE techniques. In Sec.\ref{Sec4}, the masses of
non-Goldstone type spin excitons are calculated. We give a brief
summary and conclusion in Sec.\ref{Sec5}.

\section{Dyson-Schwinger equation and dynamical mass generation}\label{Sec2}

Since the seminal work of Nambu and Jona-Lasinio, DCSB has been
investigated in the context of particle physics for nearly fifty
years \cite{Nambu_Jona-Lasinio1961, Gross_Neveu1974, Rosenstein,
Miransky, Roberts1994, Appelquist1988, Appelquist1999}. In
particular, it is one of the most prominent features of QCD.
Unfortunately, the structure of QCD is too complicated, so the
problem of DCSB in QCD has not yet been solved satisfactorily,
although there have been remarkable progress in its supersymmetric
version \cite{Witten}.

In order to gain insights into QCD, some theorists turn to QED$_3$.
Despite its simple structure, QED$_3$ shares a number of salient
features with QCD: asymptotic freedom \cite{Appelquist1995}, 
DCSB \cite{Appelquist1988}, and confinement
\cite{Polyakov1977, Maris}. Specifically, Appelquist \emph{et}
\emph{al.} found that DCSB can take place when the fermion flavor is
less than certain critical value, $N_f < N_f^c$ in
QED$_3$\cite{Appelquist1988}. Besides its relevance to particle
physics, QED$_3$ of massless Dirac fermions also has important
applications in condensed matter physics. Indeed, it is the
low-energy effective field theory for a wide class of planar
strongly correlated electron systems, especially high temperature
cuprate superconductor \cite{LeeWen}. The DCSB in QED$_3$ is
interpreted as the formation of two-dimensional Heisenberg
antiferromagnetic order, which is the ground state of undoped cuprates
\cite{LeeWen}.

In the case of graphene, the Coulomb interaction between Dirac
fermions plays an essential role since it is poorly screened. There
is a close similarity between the low-energy continuum theory of
graphene and QED$_3$. It is therefore not surprising that Coulomb
interaction can lead to DCSB. However, there is also important
difference: the Coulomb potential is non-relativistic and contains
only the temporal component of gauge field. Furthermore, although
the massless Dirac fermions are two-dimensional, the electromagnetic
field propagates in three spatial dimensions.

Following Ref.\cite{Gorbar_Gusynin2002}, we describe the Dirac
fermion in graphene by reducing QED$_4$ to (2+1)-dimension. The
effective action has the form
\begin {eqnarray}\label{action}
S &=& \int dt d^{2}r \bar{\psi}_{s}(t,\mathbf{r})(i\gamma^{0}
\partial_{t}-iv_F\gamma^{i}\partial_{i}-m_{s})\psi_{s}(t,\mathbf{r})
\nonumber\\
&& -\frac{1}{2}\int dt dt' d^{2}r d^{2}r'
\bar{\psi}_{s}(t,\mathbf{r})\gamma^{0}\psi_{s}(t,\mathbf{r})
U_{0}(t-t',|\mathbf{r}-\mathbf{r'}|)\bar{\psi}_{s'}(t',\mathbf{r}')
\gamma^{0}\psi_{s'}(t',\mathbf{r}'),
\end{eqnarray}
where the Fermi velocity $v_F=c/300$ and subscripts $s,s'=1,2 (\mathrm{or}\uparrow,\downarrow)$ are
spin indices. The spinor field is
$\psi = (A_K,B_K,B_{K'},A_{K'})$ with sublattice indices $A,B$ and
momentum valley indices $K,K'$. The $\gamma$ matrices are defined as
\begin{displaymath}
\gamma_0=
\left(\begin{array}{cc}
 0 & 1_2 \\
1_2 & 0
\end{array}\right),
\gamma_i = \left(\begin{array}{cc}
 0 & -\sigma_i \\
\sigma_i & 0
\end{array}\right),
\gamma_3 = \left(\begin{array}{cc}
 0 & -\sigma_3 \\
\sigma_3 & 0
\end{array}\right),
\gamma_5=
\left(\begin{array}{cc}
1_2 & 0 \\
 0 & -1_2
\end{array}\right),
\end{displaymath}
which are used by \cite{Gusynin2007} and \cite{Ryu_Chamon2009},
and obey the Clifford algebra $\{\gamma_\mu,\gamma_\nu\} =
2g_{\mu\nu}$ with $\mu,\nu=0,1,2$. Both $\gamma_3$ and $\gamma_5$
anti-commute with $\gamma_{\mu}$. Note that different
choices of the $\gamma$ matrices do not change the final results
obtained in this section. The bare Coulomb potential $ U_{0} (t,r)$
between Dirac fermions is given by
\begin {equation}\label{bareCoulomb}
U_{0} (t,r)=\frac{e^{2}\delta(t)}{\epsilon r},
\end{equation}
where $\epsilon$ is the dielectric constant. In graphene, it is
convenient to define a fine structure constant as $\alpha =
e^{2}/\epsilon \hbar v_{F}$. It takes different values when graphene is
placed on different substrates. For graphene on substrate SiO$_2$,
$\alpha \approx 0.8$; for graphene in vacuum (suspended), $\alpha =
2.2$ \cite{Son2007, Castro-Neto2009}.

When the Dirac fermions are massless, $m_s = 0$, the Lagrangian (1)
has an exact continuous chiral symmetry, $\psi \rightarrow
e^{i\theta\gamma_{3,5}}\psi$. However, this symmetry is not
respected by the mass term $\bar{\psi}\psi$. Besides, when a finite fermion
mass is generated by formation of excitonic pairs due to Coulomb
interaction, this symmetry will be dynamically broken. Using the DS
equation approach, it is argued that a sufficiently strong
long-range Coulomb interaction can induce a dynamical fermion mass,
$\langle \bar{\psi}\psi \rangle \neq 0$, thus leading to quantum
phase transition from semimetal to excitonic insulator
\cite{Khveshch2004, Gorbar_Gusynin2002, Liu2009, Gamayun2010}.

When the Dirac fermions have small bare mass $m_s$ with $s=1,2$,
the Lagrangian respects only an approximate chiral symmetry. We will
solve the corresponding gap equation in the presence of a small bare
fermion mass and calculate the vacuum chiral condensate $\langle
\mathrm{vac}|\bar{\psi}_{s}\psi_{s'}|\mathrm{vac}\rangle$, which is
essential for the later computation of exciton mass in the next
section. Due to the above definition of gamma matrices, the mass
term $\bar{\psi}\psi$ corresponds to Kekule distortion
\cite{Gusynin2007,Dillenschneider2008}. More generally,
Kekule distortion is formulated by 
$(\mathrm{Re}\delta)\bar{\psi}\psi+(\mathrm{Im}\delta)\bar{\psi}i\gamma_5\psi$ \cite{Gusynin2007},
but it can be transformed into $m\bar{\psi}\psi$ with 
$m=\sqrt{(\mathrm{Re}\delta)^2+(\mathrm{Im}\delta)^2}=|\delta|$ by 
absorbing the phase into the 4-spinor $\psi$ \cite{Hou_Chamon2007}.
We should also note that the physical origin of bare mass do not
affect the solution of gap equation and the magnitude of the
corresponding mass generation only depends on the magnitude of
the bare mass.
For simplicity, we are mainly interested in the case where
$m_1$ and $m_2$ are nearly the same, i.e., $|m_1-m_2| \ll m$ with $m
= (m_1+m_2)/2$.

The DS equation for Dirac fermion propagator of $\psi_1$ 
reads \cite{Gamayun2010}
\begin{equation}\label{DSequ}
S^{-1}(p_0,\mathbf{p})=S^{-1}_{0}(p_0,\mathbf{p}) -ie^{2}\int
\frac{dk_{0}}{(2\pi)}\frac{d^{2}k}{(2\pi)^{2}}
V(p_0-k_{0},\mathbf{p}-\mathbf{k})
\gamma^{0}S(k_{0},\mathbf{k})\gamma^{0},
\end{equation}
where the bare propagator is
\begin{equation}
S_{0}(p_0,\mathbf{p}) = \frac{1}{p_0\gamma^{0} -
\mathbf{p}\mathbf{\gamma} - m_1},
\end{equation}
and the full fermion propagator has the form
\begin{equation}\label{FermiPropagator}
S(p_0,\mathbf{p}) =
\frac{1}{p_0\gamma^{0}-\mathbf{p}\mathbf{\gamma}-\Delta_1(p_0,\mathbf{p})}.
\end{equation}
Hereafter we use the unit $v_F=1$.
Following previous works \cite{Khveshch2004, Gorbar_Gusynin2002}, we
adopt the instantaneous approximation and replace the gap function
$\Delta_1(p_0,\mathbf{p})$ by $\Delta_1(0,\mathbf{p})$. Then we have
\begin{equation}\label{gap1}
\Delta_1(\mathbf{p}) = m_1+\frac{\alpha}{2} \int \frac{d^{2}k}{2\pi}
J(\mathbf{p},\mathbf{k})
\frac{\Delta_1(\mathbf{k})}{\sqrt{\mathbf{k}^{2}+\Delta_1(\mathbf{k})^{2}}}
\end{equation}
with the kernel function
\begin{equation}
J(\mathbf{p},\mathbf{k}) =
\frac{\beta(\mathbf{p}-\mathbf{k})}{|\mathbf{p}-\mathbf{k}|},
\end{equation}
where
\begin{equation}\label{kernel_beta}
\beta(\mathbf{q}) =
\frac{1}{1+\frac{N_f}{2}\alpha\left[\frac{2m_1}{|\mathbf{q}|} +
\frac{\mathbf{q}^{2} - 4m_{1}^{2}}{\mathbf{q}^{2}}
\arctan\left(\frac{|\mathbf{q}|}{2m_1}\right)\right]}.
\end{equation}
$N_f$ is the number of spin flavor and will be taken to be $2$ in
this paper. We approximate the kernel by its asymptotic value at $p
\ll k$ and $p \gg k$, so that
\begin{equation}\label{kernel_asy}
J(p,k) = \theta(p-k)\frac{\beta(\mathbf{p})}{p} +
\theta(k-p)\frac{\beta(\mathbf{k})}{k}
\end{equation}
Therefore, the gap equation Eq.(\ref{gap1}) is written in the form
\begin{equation}\label{gap2}
\Delta_1(p) = m_1 + \frac{\alpha}{\pi}\int^{\Lambda}_0 dk
J(p,k)\frac{k\Delta_1(k)}{\sqrt{k^{2}+\Delta_1(k)^{2}}},
\end{equation}
where $\Lambda$ is upper momentum cut-off with the order of the
inverse of the lattice constant. Here we choose its value as
10eV.

In the present work, we solve the integral equation numerically.
From $\Delta(p)$, the chiral condensate
$\langle\mathrm{vac}|\bar{\psi}_{s}\psi_{s'}|\mathrm{vac}\rangle$
can be evaluated by its definition\cite{Gamayun2010}
\begin{eqnarray}\label{condensate}
\langle\mathrm{vac}|\bar{\psi}_{s}\psi_{s'}|\mathrm{vac}\rangle &=&
- \mathrm{tr}\lim_{x\rightarrow 0}
\langle\mathrm{vac}|T\psi_{s'}(x)\bar{\psi}_{s}(0)|\mathrm{vac}\rangle
= -\mathrm{tr} \int
\frac{dp_0}{2\pi}\frac{d^{2}p}{(2\pi)^{2}}\frac{i}{\ps -
\Delta_{s}}\delta_{ss'} \nonumber \\
&=& -\frac{1}{\pi}\int_0^{\Lambda}\frac{\Delta_{s}(p) p
dp}{\sqrt{p^{2}+\Delta_{s}(p)^{2}}}\delta_{ss'}.
\end{eqnarray}

The numerical results for chiral condensate
$\langle\mathrm{vac}|\bar{\psi}_{s}\psi_{s'}|\mathrm{vac}\rangle =
\delta_{ss'}\langle\bar{\psi}\psi\rangle$ and dynamical fermion mass
at zero momentum $\Delta(0)$ are present in Table 1 for the case of
zero bare mass $m=0$ and in Table 2 for the case of small bare mass
$m =1.00\times 10^{-7}\Lambda$. Comparing these two tables, it is
easy to see that the small bare mass of Dirac fermion leads to a
substantial enhancement of chiral condensate. When interaction
strength $\alpha = 5$, the dynamical mass $\Delta(0)$ in Table 2 is
nearly twenty times larger than the value of dynamical mass in Table
1, while the chiral condensate is about one hundred times larger.
When there is no bare fermion mass, there is no excitonic pairing
and the Dirac fermions remain massless for $\alpha \leq 2.4$. The
critical strength $\alpha_c$ of the Coulomb interaction lies in the
interval of $(2.4,2.5)$, which is a little larger than the result of
\cite{Gamayun2010}. However, when the Dirac fermions have a small
bare mass, excitonic pairing and DCSB can take place for any finite
value of $\alpha$. For graphene placed on SiO$_2$ with $\alpha
\approx 0.8$ \cite{Son2007, Castro-Neto2009}, the dynamical fermion
mass $\Delta(0) = 2.15 \times 10^{-6}\Lambda$ comes mainly from the
formation of excitonic pairs since it is much larger than the bare
mass $1.00\times 10^{-7} \Lambda$.

In summary, the small bare fermion mass has two effects: it reduces
the critical Coulomb interaction strength $\alpha_c$ and enhances
the magnitude of dynamical fermion mass $\Delta(0)$. These effects
are important from both experimental and technical points of view.
On the one hand, the fermion gap can be detected unambiguously in
experiments only when it is sufficiently large
\cite{Castro-Neto2009}. On the other hand, the graphene material
with a large gap will have more technical advantages than that
with a negligible gap \cite{Castro-Neto2009}.

We would like to emphasize that, although the bare fermion mass can
catalyse the generation of dynamical fermion mass, the latter is
physically different from the bare mass. In this work, the bare mass
is assumed to be generated by several possible mechanisms and small
in quantity. It can be non-zero even when the Coulomb interaction
between Dirac fermions is completely ignored. However, the dynamical
fermion mass originates from the formation of chiral vacuum condensate
driven by Coulomb interaction and thus is a typical effect of strong
correlation between fermions.

It is also interesting to compare the excitonic pair formation with
the Cooper pair formation in the BCS theory of superconductivity.
Historically, the proposal of dynamical chiral symmetry breaking of
Nambu and Jona-Lasinio was motivated by the BCS theory. However,
there are essential differences between them. An excitonic pair is
composed of a particle and a hole, and thus is neutral. On the
contrary, a Cooper pair is composed of two electrons, and thus
carries negative charge $-2e$. Moreover, the formation of excitonic
pairs breaks chiral symmetry and leads to insulating behavior, whereas
the formation of Cooper pairs break local gauge symmetry and lead to
superconductivity.

The DS equation with a Kekule-distortion induced fermion mass was
studied previously in Ref.~\cite{Dillenschneider2008}, which
concludes that the dynamical mass generation due to interaction is
independent of the homogeneous Kekule distortion. In
Ref.~\cite{Dillenschneider2008}, an important claim was that the gap
equation is dominated by the large momentum regime. However, it is
known that dynamical chiral symmetry breaking is a non-perturbative,
low-energy phenomenon. This phenomenon can happen only when the
interaction between fermions is weak at high energy/momentum regime
and strong at low energy/momentum regime (i.e., asymptotic freedom).
Therefore, the processes with small energy/momentum transfer should
play dominant role in the formation of excitonic pairing. According
to our numerical computation, the dynamical mass generation is
indeed significantly affected by the presence of homogeneous Kekule
distortion whose catalytic effect can be readily seen from the
comparison of Table I and Table II.

\begin{table}[pth]
\caption{Numerical results of chiral condensate without bare mass.
Here, the chiral condensate is $\rho=
-\langle\bar{\psi}\psi\rangle/\Lambda^{2}$
\label{masslessCondensate}.}
\begin{center}
\begin{tabular}{|c|c|c|c|c|c|}
\hline
 $\alpha$                      & 2.4 & 2.5      & 5      & 10         & $\infty $
\\\hline

$\rho$                         & 0 & 1.34(-31) & 1.84(-9) & 2.22(-7)    & 4.57(-6)
\\\hline

$\Delta(0) / \Lambda $          & 0 & 2.67(-21)  & 1.70(-6) & 4.37(-5) & 3.45(-4)
\\\hline
\end{tabular}
\end{center}
\end{table}

\begin{table}[pth]
\caption{Numerical results of chiral condensate with baremass $m=10^{-7}\Lambda$.
\label{masssiveCondensate} }
\begin{center}
\begin{tabular}{|c|c|c|c|c|c|c|c|c|c|c|}
\hline
$\alpha$                 & 0       &0.8     & 1       & 2       & 2.2     &2.4     & 2.5      &5        & 10      &$\infty$  \\\hline
$\rho$                   & 3.20(-8)&5.40(-8)&5.91(-8) &8.57(-8) &9.17(-8) &9.78(-8) & 1.01(-7) &2.13(-7)&6.37(-7)  & 5.12(-6) \\\hline
$\Delta(0)/\Lambda$      & 1.00(-7) &2.15(-6)& 3.00(-6)& 8.37(-6)& 9.62(-6)&1.09(-5)&1.16(-5)& 3.12(-5)& 8.19(-5)& 3.67(-4)       \\\hline

\end{tabular}
\end{center}
\end{table}

In the calculation presented below, we will need the normal-ordered
chiral condensate $\langle
\mathrm{vac}|:\bar{\psi}(0)\psi(0):|\mathrm{vac}\rangle = \langle
\mathrm{vac}|\bar{\psi}(0)\psi(0)|\mathrm{vac}\rangle -
\langle\Omega|\bar{\psi}(0)\psi(0)|\Omega\rangle$. Quite different
from the non-perturbative symmetry-broken vacuum state
$|\mathrm{vac}\rangle$, $|\Omega \rangle$ is the perturbative vacuum
state that is chiral symmetric \cite{Zong2003}. As a function of
$\alpha$, the value of $\langle
\mathrm{vac}|:\bar{\psi}(0)\psi(0):|\mathrm{vac}\rangle$ will be
specified in Sec.\ref{Sec3}.

\section{SVZ sum rule analysis of Goldstone type excitons}\label{Sec3}

Since the Dirac fermions have a small bare mass, the Lagrangian of
graphene respects an approximate continuous chiral symmetry. As
emphasized by Weinberg \cite{Weinberg}, when an approximate
continuous symmetry is broken, the Goldstone bosons are no longer
massless. Instead, these bosons are massive. In the context of
graphene, the Goldstone type excitons induced by dynamical breaking
of approximate chiral symmetry have finite masses. In this section,
we calculate the masses of these excitons and compare them with the
dynamical fermion mass.

An exciton is a boson composed of a Dirac particle and a Dirac hole
(i.e., an anti-fermion in the terminology of particle physics). It
can be described by a composite operator of spinor field and its
conjugate. For graphene, when chiral condensate occurs $ \langle
\bar{\psi}\psi \rangle \neq 0$, the chiral symmetry U(4) is broken
down to $U(2)\times U(2)$ and the number of broken generators is
$8$. The $8$ corresponding Goldstone bosons can be described by
$\bar{\psi} \gamma_3 \otimes \sigma_\mu \psi$ and $\bar{\psi}
i\gamma_5 \otimes \sigma_\mu \psi$ \cite{Appelquist1988, Burden1992},
where $\sigma_0$ is unit matrix. The modes of $\bar{\psi}i\gamma_5
\otimes\sigma_{\mu}\psi$ are bond-density-waves that mix the $K$ and
$K'$ points, similar to the Kekule distortion mode
$\langle\bar{\psi}\psi\rangle$ \cite{Ryu_Chamon2009,
Hou_Chamon2007}. $\bar{\psi}\gamma_3 \otimes\sigma_{\mu}\psi$ is
related to the CDW or staggered SDW excitations \cite{Gusynin2007}.

In this section, we will calculate the masses of the lowest-energy
excitons corresponding to $\phi_0 = \bar{\psi}\gamma_3
\otimes\sigma_{0}\psi$ and $\phi_i = \bar{\psi}\gamma_3
\otimes\sigma_i \psi$ using the SVZ sum rule method, which is
analogous to the procedure of computing pion mass in particle
physics \cite{SVZ1979b}. This procedure can also be applied to the
excitons associated with $\bar{\psi}i\gamma_5
\otimes\sigma_{\mu}\psi$ and the results are the same, so we will
not discuss the case of $\bar{\psi}i\gamma_5
\otimes\sigma_{\mu}\psi$. The basic idea of SVZ sum rule is to
compute one particular physical quantity in two different ways and
then extract important information of some parameter (such as the
mass of a Goldstone boson) by equating the expressions obtained by
different ways. In the present problem, the physical quantities to be
computed are the correlation functions of composite fields $\phi_0$
and $\phi_i$. To apply the SVZ sum rule technique, OPE method will
be used to separate the perturbative and non-perturbative
contributions to the correlation function. The perturbative
contributions are included in the so-called Wilson's coefficients
and can be calculated perturbatively. The non-perturbative
contributions are embodied as chiral condensates, which can be
obtained from experimental data or calculated by some
non-perturbative methods. In our case, the condensates are
calculated by means of DS equation method.

We first consider the field $\phi_0$. Its correlation function is
defined as
\begin{equation}\label{Pi_correlator}
\Pi(q) = i\int d^{3}x e^{iqx} \langle \mathrm{vac}|
T\phi_{0}(x)\phi^{\dag}_{0}(0)|\mathrm{vac}\rangle,
\end{equation}
where $|\mathrm{vac}\rangle$ is the non-perturbative vacuum state in
the chiral symmetry breaking phase. Substituting the 4 components of
$\psi$, we have
\begin{equation}\label{CDW}
\phi_0 \sim \sum_s A_s^{+}(K)A_s(K)+A_s^{+}(K')A_s(K') -
B_s^{+}(K)B_s(K)-B_s^{+}(K')B_s(K') = N_A-N_B,
\end{equation}
which corresponds to the CDW excitation. In order to get information
of the excitation, the correlate function will be calculated in two
different ways: phenomenologically and theoretically. On the
phenomenological side, the correlation function is related to
physical observables, while on the theoretical side the same function
is expressed in terms of fundamental parameters such as $m$ and
$\alpha$ which are treated as known numbers. Equating the results
obtained in these two ways, we can obtain the expression for exciton
masses in terms of fundamental parameters.

To perform the phenomenological computation, we insert a complete
set of physical states, $1 = \sum_{n}|n,p_n\rangle d\tau_n\langle
n,p_n|$, between the two operators in Eq.(\ref{Pi_correlator}).
Here, $p_n$ is the momentum of the intermediate state and $d\tau_n =
d^3 p_n\delta(p^2_n-M^2_n) \theta(p^0_n)/(2\pi)^2$ denotes the
integration over phase-space \cite{Colangelo_Khodjamirian2000}. The
index $n$ can take both discrete and continuous values, which
implies that both discrete bound states and continuous states are
included. After integrating out space and momentum coordinates, we
obtain
\begin{equation}\label{PiIm}
\mathrm{Im}\Pi(q)=\sum_n\pi \delta(q^2-M_{n}^2)F_n,
\end{equation}
where $F_n=|\langle n,q|\phi_0|\mathrm{vac}\rangle|^2$. $M_{n}$ is
the mass of the intermediate state $|n\rangle$ or the static energy
in its rest reference system. For later convenience, we now
introduce a function $\widetilde{\Pi}(q^2)$, defined by
\begin{equation}\label{PiTilde}
\Pi(q)=q^2\widetilde{\Pi}(q^2),
\end{equation}
and then have
\begin{equation}\label{PiTildeIm}
\mathrm{Im}\widetilde{\Pi}(q^2) = \pi
f_{\varphi_0}\delta(q^2-m_{\varphi_0}^2)+{\sum_n\hspace{-0.46cm}\int}\pi
f_{n}\delta(q^2-M_{n}^2)
\end{equation}
with $F_n=M^2_n f_n$. Here we have isolated the contribution of the
lowest-lying state $\varphi_0$. The mass of $\varphi_0$,
$m_{\varphi_0}$, gives the position of the corresponding resonance.
In the context of particle physics, $f_{\varphi_0}$ is called decay
constant of the bound state. In the context of graphene, it is
related to the strength of exciton resonance. The second term of the
righthand side of Eq.(\ref{PiTildeIm}) includes both discrete and
continuous states. Using the dispersion relation
\begin{equation}\label{dispersion}
\widetilde{\Pi}(q^2) = \frac{1}{\pi}\int_0^{+\infty} ds
\frac{\mathrm{Im}\widetilde{\Pi}(s)}{s-q^2-i\epsilon},
\end{equation}
we obtain
\begin{equation}\label{PiPheno}
\widetilde{\Pi}(q^2) = \frac{f_{\varphi_0}}{m_{\varphi_0}^2-q^2} +
{\sum_n\hspace{-0.46cm}\int}\frac{f_{n}}{M_n^2-q^2},
\end{equation}
where the first term is the contribution of the lowest-lying state
and the second term is the contribution of higher excited states.

In order to identify the mass $m_{\varphi_0}$ of the lowest-lying
exciton state corresponding to $\bar{\psi}\gamma_3\psi$, we need to
perform the theoretical analysis starting from the same correlation
function. The correlation function can be calculated by means of OPE
method, which is able to account for the non-perturbative effects
due to chiral-symmetry broken vacuum. We first write down the
following expression
\begin{eqnarray}\label{PiOPE}
i\int d^{3}x e^{iqx} T\phi_{0}(x)\phi^{\dag}_{0}(0) = C_0(q) +
C_{2,1}(q):\bar{\psi}_1\psi_1(0): +
C_{2,2}(q):\bar{\psi}_2\psi_2(0):+
C_4(q):\bar{\psi}\Gamma\psi\bar{\psi}\Gamma'\psi(0): + ...
\end{eqnarray}
The Wilson's coefficients $C_n$'s contain the perturbative
contributions, while the operators multiplying $C_n$'s contain the
non-perturbative contributions. From dimension analysis, we know
that the mass dimension of the left hand side of Eq.(\ref{PiOPE}) is
one, so each term in the right hand side should have the same
dimension. Since the operators appearing in the right hand side are
all local operators with increasing mass dimensions, the mass
dimension of $C_n$ should decrease with $n$. This indicates that
$C_n$'s contain increasing powers of $1/q^2$ as $n$ grows. After
taking vacuum expectation value, the contributions of higher terms
are suppressed by $\langle :\bar{\psi}\psi: \rangle/q^2$ for large
$q^2$. In the present work, we are mainly interested in the region
$-q^2 \gg |\langle:\bar{\psi}\psi:\rangle|$ in Eq.(\ref{PiOPE}) and
thus can keep only the first three terms. Generally speaking, $C_n$
can be obtained by sandwiching the two sides of Eq.(\ref{PiOPE})
with a pair of states and comparing the results from both sides, as
shown by Braateen \cite{Braaten2008}. Here for $C_0$, we sandwich
Eq.(\ref{PiOPE}) by perturbative vacuum state $|\Omega\rangle$,
which eliminates the contributions of higher terms. Therefore $C_0$
will be calculated by evaluating the Feynman diagram of a fermion
loop. For $C_{2,1}$ and $C_{2,2}$, following the works of
\cite{Colangelo_Khodjamirian2000, Pascual1984}, we sandwich
Eq.(\ref{PiOPE}) by non-perturbative vacuum state
$|\mathrm{vac}\rangle$ and apply Wick's theorem to the left-hand
side. Then a pair of fermionic operators are contracted. After
calculating tree-level Feynman diagram and matching the two sides,
we will get $C_{2,1}$ and $C_{2,2}$. Furthermore, perturbative
calculations for Wilson's coefficients are performed in the
framework of $1/N_f$ expansion and only leading term $C^{(0)}_n$'s
are kept in the present work.

Sandwiching Eq.(\ref{PiOPE}) by the physical vacuum state, one
obtains
\begin{equation}\label{PiOPE_vev1}
\Pi(q) = C_0^{(0)}(q)+ C^{(0)}_{2,1}(q)\langle:\bar{\psi}_1\psi_1:\rangle +
C^{(0)}_{2,2}(q)\langle:\bar{\psi}_2\psi_2:\rangle.
\end{equation}
Similar to the relation between $\Pi$ and $\widetilde{\Pi}$, we
introduce $\widetilde{C}_n^{(0)}$'s satisfying
\begin{equation}\label{PiCTilde}
C_n^{(0)}(q)=q^2\widetilde{C}_n^{(0)}(q^2),
\end{equation}
and then have
\begin{equation}\label{PiOPE_vev2}
\widetilde{\Pi}(q^2)=\widetilde{C}_0^{(0)}(q^2) +
\widetilde{C}_{2,1}^{(0)}(q^2)\langle:\bar{\psi_1}\psi_1:\rangle +
\widetilde{C}_{2,2}^{(0)}(q^2)\langle:\bar{\psi_2}\psi_2:\rangle.
\end{equation}
Since $\mathrm{Im}C_0^{(0)}$ can be obtained from evaluating the imaginary part
of a fermionic loop diagram, we obtain
\begin{equation}\label{PiImC0}
\mathrm{Im}\widetilde{C}_0^{(0)}(q^2) =
\frac{1}{4\sqrt{q^2}}\theta(q^2-4m_1^2)+\frac{1}{4\sqrt{q^2}}\theta(q^2-4m_2^2).
\end{equation}
$C^{(0)}_{2,i}$ (i=1,2) can be obtained by
directly computing the tree diagrams, with the expression
\begin{equation}\label{PiC2-1}
C_{2,i}^{(0)}(q^2) = \frac{4m_i q^2}{3(q^2-m_i^2)^2} \simeq
\frac{4m_i }{3q^2},
\end{equation}
in the region $Q^2 = -q^2 \gg m^2$.

Equating the right hand sides of Eq.(\ref{PiPheno}) and
Eq.(\ref{PiOPE_vev2}) and substituting $C_n$'s, we obtain
\begin{eqnarray}\label{Pisumrule1}
\frac{f_{\varphi_0}}{Q^2+m_{\varphi_0}^2}+{\sum_n\hspace{-0.46cm}\int}
\frac{f_{n}}{Q^2+M_n^2} = \frac{1}{\pi}\int ds
\frac{\mathrm{Im}\widetilde{C}_0^{(0)}(s)}{s+Q^2-i\epsilon} +
\frac{4m_1\langle:\bar{\psi}_1\psi_1:\rangle}{3(Q^2)^2} +
\frac{4m_2\langle:\bar{\psi}_2\psi_2:\rangle}{3(Q^2)^2} +
\mathcal{O}(\frac{1}{Q^6}).
\end{eqnarray}
In order to extract the information of the lowest-lying state, it is
helpful to introduce the Borel transformation
\begin{equation}\label{Borel}
\mathcal{B}_{M^2} = -\lim_{Q^2=n M^2\rightarrow \infty}
\frac{(-Q^2)^{n+1}}{n!}(\frac{d}{dQ^2})^n,
\end{equation}
which can suppress the contribution from more massive states on the
left hand side of Eq.(\ref{Pisumrule1}). Now we can obtain the sum
rule
\begin{equation}\label{Pisumrule2}
f_{\varphi_0} e^{-m^2_{\varphi_0}/M^2}+{\sum_n\hspace{-0.46cm}\int}
f_n e^{-M_n^2/M^2} = \frac{1}{\pi}\int ds
\mathrm{Im}\widetilde{C}_0^{(0)}(s)e^{-s/M^2} - \frac{2X}{3 M^2} +
\mathcal{O}(\frac{1}{M^4}),
\end{equation}
where
\begin{equation}\label{X}
X = -2m_1\langle:\bar{\psi}_1\psi_1:\rangle -
2m_2\langle:\bar{\psi}_2\psi_2:\rangle.
\end{equation}
The contribution from higher states is suppressed exponentially and
seems to be unimportant. However, these higher states include
continuous ones, so they can not be simply neglected. We assume the
continuous states begin at $2\Delta(p=0)$ with $\Delta(0)$ being the
dynamical fermion mass, and subtract their contribution from both
sides of Eq.(\ref{Pisumrule2}) which is an analogue to the
quark-hadron duality in particle physics \cite{Shifman1998,
Colangelo_Khodjamirian2000, Shifman2000, Hofmann2004}. The resultant
expression is
\begin{equation}\label{Pisumrule_sbtrct}
\frac{1}{\pi}\int_0^{(2\Delta)^2} ds
\mathrm{Im}\widetilde{\Pi}(s)e^{-s/M^2} =
\frac{1}{\pi}\int_0^{(2\Delta)^2} ds
\mathrm{Im}\widetilde{C}_0^{(0)}(s)e^{-s/M^2}-\frac{2X}{3
M^2}+\mathcal{O}(\frac{1}{M^4}),
\end{equation}
In this expression, $X$ contains the normal-ordered condensates
which were not specified in Sec.II. Here we defined them as
\begin{eqnarray}\label{condensate_sbtrct}
\langle \mathrm{vac}|:\bar{\psi}_1\psi_1:|\mathrm{vac}\rangle =
-\frac{1}{\pi}\int_0^{2\Delta_1 (0)}\frac{\Delta_1(p) p
dp}{\sqrt{p^{2}+\Delta_1(p)^{2}}},
\end{eqnarray}
and $\langle:\bar{\psi}_2\psi_2:\rangle$ can be defined similarly.
From the numerical results of $\Delta(p)$ obtained Sec.II, the value
of $\langle:\bar{\psi}\psi:\rangle$ can be computed by numerical
integration. Then we obtain the sum rule formula
\begin{equation}\label{Pisumrule_3}
f_{\varphi_0} e^{-m^2_{\varphi_0}/M^2}+ ...
=\frac{1}{2\pi}\int_{2m_1}^{2\Delta_1}e^{-u^2/M^2}du +
\frac{4m_1\langle:\bar{\psi_1} \psi_1:\rangle}{3 M^2} +
(1\leftrightarrow 2)+\mathcal{O}(\frac{1}{M^4}).
\end{equation}
where $...$ is the contribution from the discrete higher states and
is neglected altogether. Differentiating the above expression with
respect to $\eta=1/M^2$ and introducing the ratio between the
derivative and Eq.(\ref{Pisumrule_3}), we eventually obtain the sum
rule for the mass of exciton $\varphi_0$:
\begin{equation}\label{Pisumrule4}
m^2_{\varphi_0}= \frac{\int_{2m_1}^{2\Delta_1}u^2 e^{-u^2\eta}du -
\frac{8\pi}{3}m_1\langle:\bar{\psi}_1\psi_1:\rangle +
(1\leftrightarrow 2)}{\int_{2m_1}^{2\Delta_1}e^{-u^2\eta}du +
\frac{8\pi}{3}m_1\langle:\bar{\psi}_1\psi_1:\rangle\eta +
(1\leftrightarrow 2)}.
\end{equation}
Besides, the sum rule formula for $f_{\varphi_0}$ is
\begin{equation}\label{Pisumrule_5}
f_{\varphi_0} =
e^{m^2_{\varphi_0}\eta}\left[\frac{1}{2\pi}\int_{2m_1}^{2\Delta_1}e^{-u^2\eta}du
+ \frac{4m_1\langle:\bar{\psi_1}\psi_1:\rangle}{3}\eta
+(1\leftrightarrow 2)\right].
\end{equation}

We next consider the correlation function of
$\phi_1 = \bar{\psi}_{1}\gamma_3\psi_{2}
\sim A^{\dagger}_{\uparrow}A_{\downarrow}-B^{\dagger}_{\uparrow}B_{\downarrow}$.
Analogously, we define
\begin{equation}\label{correlator}
\Pi_1(q)=i\int d^{3}x e^{iqx} \langle \mathrm{vac}|
T\phi_{1}(x)\phi^{\dag}_{1}(0)|\mathrm{vac}\rangle,
\end{equation}
which is the counterpart of the transverse component of staggered
spin susceptibility investigated in the context of high temperature
cuprate superconductors \cite{Rantner_Wen2002}. Similar OPE is
applied to $\Pi_1(q)$, and the results for the corresponding
coefficients are
\begin{equation}\label{Pi_1_ImC0}
\mathrm{Im}\widetilde{C}_0^{(0)}(q^2) =
\frac{1}{4\sqrt{q^2}}\theta(q^2-4m^2)
\end{equation}
with $m=(m_1+m_2)/2$. The resulting sum rule formula after Borel
transformation is
\begin{equation}\label{Pi_1_sumrule1}
f_{\varphi_1} e^{-m^2_{\varphi_1}/M^2}+ ... =
\frac{1}{2\pi}\int_{2m}^{2\Delta}e^{-u^2/M^2}du +
\left[\frac{(3m_2-m_1)\langle:\bar{\psi}_1 \psi_1:\rangle}{3 M^2} +
(1\leftrightarrow 2)\right] + \mathcal{O}(\frac{1}{M^4}).
\end{equation}
Differentiate it with respect to $\eta=1/M^2$ then one gets the sum
rule formula for the mass of exciton $\varphi_1$:
\begin{equation}\label{Pi_1_sumrule2}
m^2_{\varphi_1} = \frac{\int_{2m}^{2\Delta}u^2 e^{-u^2\eta}du -
[\frac{2\pi}{3}(3m_2-m_1)\langle:\bar{\psi}_1\psi_1:\rangle +
(1\leftrightarrow 2)]}{\int_{2m}^{2\Delta}e^{-u^2\eta}du +
[\frac{2\pi}{3}(3m_2-m_1)\langle:\bar{\psi}_1\psi_1:\rangle\eta +
(1\leftrightarrow 2)]}
\end{equation}
with $2m=m_1+m_2$ and $2\Delta=\Delta_1(0)+\Delta_2(0)$.

The sum rule formulae for the masses of excitons $\phi_2$ and
$\phi_3$ can be derived similarly. The formula for $m_{\varphi_2}$
is exactly the same as $m_{\varphi_1}$, Equ.(\ref{Pi_1_sumrule2}).
For $\phi_3$, one considers the following two-point correlation
function
\begin{equation}\label{correlator3_3}
\Pi_3(q)=i\int d^{3}x e^{iqx} \langle \mathrm{vac}|
T\phi_{3}(x)\phi_{3}(0) |\mathrm{vac}\rangle,
\end{equation}
where $\phi_3$ is the longitudinal component of the staggered SDW,
given by
\begin{equation}\label{stagger}
\phi_3 \sim (N_{A\uparrow}-N_{A\downarrow})-(N_{B\uparrow}-N_{B\downarrow}).
\end{equation}
The sum rule formula for $m_{\varphi_3}$ is the same as
$m_{\varphi_0}$, which is presented above.

In our truncated computation, these sum rule formulae for the
exciton masses Eq.(\ref{Pisumrule4}) and Eq.(\ref{Pi_1_sumrule2})
are functions of the Borel parameter $M$ or $\eta$. In order to fix
the value $m_\varphi$ for exciton $\varphi$, we need a criterion for
the choice of $\eta$ and the corresponding $m_\varphi(\eta)$. In the
sum rule literature, the optimum estimate-value may be determined by
an extreme value, an inflection point \cite{Pascual1984,Durand1983}
or a plateau \cite{Colangelo_Khodjamirian2000} of the function. In
the present work, we choose a plateau near the extreme value of the
function to be the final output for $m_{\varphi}$ or $f_{\varphi}$.

For $m_1=m_2=1.00\times 10^{-7}\Lambda$ and $\alpha=0.8$ (SiO$_2$
case), the numerical results for the chiral condensate and mass gap
are $-\langle:\bar{\psi}_1 \psi_1:\rangle = -\langle:\bar{\psi}_2
\psi_2:\rangle = 1.70\times 10^{-12}\Lambda^{2}$ and $\Delta(0) =
2.15\times 10^{-6}\Lambda$. The cut-off scale $\Lambda$ is normally
set as $10$eV. From these quantities, the SVZ estimates for the
masses of the excitons can be made with the criterion discussed
above: $\varphi_0$, $\varphi_1$, $\varphi_2$ and $\varphi_3$ have
the same mass $0.028$meV. For $m_1 \neq m_2$, the four-fold mass
degenerate is lifted to two-fold: $\varphi_1$ and $\varphi_2$ have
the same mass, which is a little smaller than $m_{\varphi_0}$ and
$m_{\varphi_3}$. For the case of suspended graphene, $\alpha = 2.2$,
we have $-\langle:\bar{\psi} \psi:\rangle =3.33\times
10^{-11}\Lambda^{2}$, $\Delta(0)=9.62\times 10^{-6}\Lambda$. Now the
mass of $\varphi$ particles is about $0.098$meV.

It is now necessary to summarize and compare the relevant energy
scales discussed above. The fundamental energy scale in the present
problem is the ultra-violet cut-off $\Lambda=10\mathrm{eV}$ in
graphene, which is normally determined by the lattice constant. In
this paper, we assume a bare fermion mass $m = 1.00\times
10^{-7}\Lambda$, which may be generated by Kekule distortion or
other mechanisms. All other energy scales are derived by explicit
calculations. We note here that the exciton mass is smaller than
$2\Delta(0)$, which implies that the binding energy of the bound
state is negative.

\section{Spin susceptibility and non-Goldstone type excitons}\label{Sec4}

In addition to the CDW and staggered SDW excitons studied in the
last section, there is another kind of low-energy collective
excitations: spin excitons. These spin excitons are not generated due
to chiral symmetry breaking and thus are non-Goldstone type bosons.
In the absence of any (bare or dynamical) fermion mass, these spin
excitons are massless in clean graphene \cite{Baskaran_Jafari2002,
Jafari_Baskaran2005} and become massive in doped graphene
\cite{Ebrahimkhas_Jafari_Baskaran2009}. In the present problem, the
fermions have finite dynamical mass in the chiral-symmetry broken
phase, so the spin excitons are also massive. In order to study
these triplet spin-1 excitons, we turn to study the spin
susceptibilities which may be measured by inelastic neutron
scattering \cite{Jafari_Baskaran2005}. These quantities were studied
previously in an effective QED$_3$ theory of high-temperature
superconductors \cite{Seradjeh_Herbut2007}.

In this section, we will calculate the masses of spin excitons using
the methods presented in the last section. To do this, we first
define the spin operator as $S_i = \psi^{\dagger}_{a}
\sigma^{i}_{ab}\psi_b $, where $i=x,y,z$ and $\sigma^i$ are Pauli
matrices. The transverse and longitudinal spin susceptibilities are
defined as
\begin{equation}\label{correlator_spin_t}
\chi_{+-}(q) = i\int d^{3}x e^{iqx}\langle\mathrm{vac}| T
S_{+}(x)S_{-}(0)|\mathrm{vac}\rangle,
\end{equation}
and
\begin{equation}\label{correlator_spin_l}
\chi_{zz}(q) = i\int d^{3}x e^{iqx}\langle\mathrm{vac}| T
S_{z}(x)S_{z}(0)|\mathrm{vac}\rangle,
\end{equation}
respectively. Here, $|\mathrm{vac}\rangle$ is the non-perturbative
physical vacuum state of the system in the chiral-symmetry broken
phase. The quantity $\chi_{-+}$ can be similarly studied and will
not be discussed here.

The spin susceptibility will be computed by means of SVZ sum rule
method. We note that Rosenfelder applied the SVZ sum rule method to
analyze density-density correlation function and estimate the
position of plasmon resonance in a non-relativistic electron system
\cite{Rosenfelder1985}. In our case, the spin susceptibility will be
analyzed in both phenomenological and theoretical ways, analogous to
what we have done in the last section.

On the phenomenological side, following the previous work
\cite{Seradjeh_Herbut2007, Baskaran_Jafari2002, Jafari_Baskaran2005,
Ebrahimkhas_Jafari_Baskaran2009}, we assume that there is a
resonance below the spin gap. This assumption is embodied in the
expression
\begin{equation}\label{Im_chi}
\frac{1}{\pi}\mathrm{Im}\chi_l(p) = F_{\sigma}\delta(p^2 -
m^2_{\sigma})+\sum_{n}F_{n}\delta(p^2 - \mu^2_{n}) + \Theta(p^2 -
U_{\mathrm{sg}})\rho_{\sigma}(p^2),
\end{equation}
where the first $\delta$-function type resonance $\sigma$
corresponds to the so-called spin exciton. $\sigma = +,-,z$ for $l =
+-,-+,zz$ respectively. The spin gap $U_{\mathrm{sg}}$ is the lower
bound of the continuous spectrum. The second term is the
contribution from the discrete excited bound-states, whose masses
$\mu_n$'s are either between $m_{\sigma}$ and $U_{\mathrm{sg}}$ or
beyond $U_{\mathrm{sg}}$. From the dispersion relation formula, we
obtain that, for $p^2 < 0$,
\begin{eqnarray}\label{chi_pheno}
\chi_l(p^2) &=& \frac{1}{\pi}\int_0^{\infty}ds\frac{\mathrm{Im}
\chi_l(s)}{s-p^2-i\epsilon} \nonumber \\
&=& \frac{F_{\sigma}}{m^2_{\sigma}-p^2} +
\sum_n\frac{F_{n}}{\mu^2_{n}-p^2}+\int^{+\infty}_{U_{\mathrm{sg}}}
\frac{\rho_{\sigma}(s)ds}{s-p^2}.
\end{eqnarray}

To compute the same correlation function theoretically, OPE is
adopted to take into account the non-perturbative effects due to
chiral symmetry breaking. For transverse spin susceptibility, we
have
\begin{eqnarray}\label{OPE}
i\int d^{3}x e^{ipx} T S_{+}(x)S_{-}(0) =
D_0(p)+D_{2,1}(p):\bar{\psi}_1\psi_1(0): +
D_{2,2}(p):\bar{\psi}_2\psi_2(0): +
D_4(p):\bar{\psi}\Gamma\psi\bar{\psi}\Gamma'\psi(0): +...
\end{eqnarray}
As explained in Sec.III, the operators in the right hand side are
all local operators, so the contributions of this series are
suppressed by powers of $\langle:\bar{\psi}\psi:\rangle/q^2$.
Therefore, we can keep only the first three terms. Sandwiching
Eq.(\ref{OPE}) by the physical vacuum state, we have
\begin{equation}\label{OPE_vev}
\chi_{+-}(p) = D_0^{(0)}(p)+
D_{2,1}^{(0)}(p)\langle:\bar{\psi}_1\psi_1:\rangle +
D_{2,2}^{(0)}(p)\langle:\bar{\psi}_2\psi_2:\rangle + ...
\end{equation}
$D_n$ can be obtained by perturbative expansion computation and each
$D_n$ is a power series in coupling $\alpha$, therefore only the
leading terms of $D_n^{(0)}$ are important. For $D_0^{(0)}$, its
imaginary part is
\begin{equation}\label{ImD0}
\mathrm{Im}D_0^{(0)}(p^2,\mathbf{p}^2) =
\Theta(p^2-4m^2)\frac{\mathbf{p}^2}{8\sqrt{p^2}} +
\Theta(p^2-4m^2)\frac{m^2\mathbf{p}^2}{2 p^2\sqrt{p^2}}.
\end{equation}
After calculating the coefficients $D_n$'s in Eq.(\ref{OPE_vev}), we
get the truncated expression of transverse spin susceptibility for
the case $m_1 = m_2$, as follows
\begin{equation}\label{chi_theo}
\chi_{+-}(p^2,\mathbf{p}) = \frac{1}{\pi}\int_0^{\infty}ds
\frac{\mathrm{Im}D_0^{(0)}(s)}{s-p^2-i\epsilon} +
\frac{8m\langle:\bar{\psi}\psi:\rangle \mathbf{p}^2 }{3(-p^2)^2},
\end{equation}
where $\langle:\bar{\psi}\psi:\rangle =
\langle:\bar{\psi}_1\psi_1:\rangle =
\langle:\bar{\psi}_2\psi_2:\rangle$ and $-p^2 \gg m^2$. Note that
all terms with higher order of $m^2/p^2$ are ignored.

Equating the right hand sides of Eq.(\ref{chi_pheno}) and
Eq.(\ref{chi_theo}), we get a sum rule formula
\begin{equation}\label{sumrule_1}
\frac{1}{\pi}\int_0^{\infty}ds \frac{\mathrm{Im}
\chi_{+-}(s)}{s-p^2-i\epsilon} = \frac{1}{\pi}\int_0^{\infty} ds
\frac{\mathrm{Im} D_0^{(0)}(s)}{s-p^2-i\epsilon} +
\frac{8m\langle:\bar{\psi}\psi:\rangle \mathbf{p}^2 }{3(-p^2)^2}.
\end{equation}
In order to subtract the contribution from the states beyond the
spin gap in both sides of the above equation, duality is used and
the upper cutoff is replaced by $(2\Delta)^2$, so that
\begin{equation}\label{sumrule_sbtrct-1}
\frac{1}{\pi}\int_0^{(2\Delta)^2} ds \frac{\mathrm{Im}
\chi_{+-}(s)}{s-p^2-i\epsilon} = \frac{1}{\pi}\int_0^{(2\Delta)^2}
ds\frac{\mathrm{Im}D_0^{(0)}(s)}{s-p^2-i\epsilon} +
\frac{8m\langle:\bar{\psi}\psi:\rangle \mathbf{p}^2 }{3(-p^2)^2}.
\end{equation}
In this expression, the value of $\langle:\bar{\psi}\psi:\rangle$
was given by Eq.(\ref{condensate_sbtrct}). Now we can obtain the
following sum rule formula
\begin{equation}\label{sumrule_sbtrct-2}
\frac{F_{+}}{m^2_{+}-p^2}+... = \frac{1}{\pi}\int_0^{(2\Delta)^2} ds
\frac{\mathrm{Im}D_0^{(0)}(s)}{s-p^2-i\epsilon} -
\frac{8m\mathbf{p}^2}{3(-p^2)^2}\frac{1}{\pi}
\int_0^{(2\Delta)^2}\frac{\Delta(s)ds}{\sqrt{s+\Delta(s)^{2}}},
\end{equation}
where ... stands for the contribution from the discrete excited
bound-states. Furthermore, in order to extract the information of
the lowest-lying state, Borel transformation can be used to suppress
the contribution from excited states in the left hand side of
Equ.(\ref{sumrule_sbtrct-2}), leading to
\begin{equation}\label{sumrule_2}
F_{+} e^{-m^2_{+}/M^2}+ \sum F_{n} e^{-\mu^2_{n}/M^2} =
\frac{\mathbf{p}^2}{4\pi}\int_{2m}^{2\Delta}e^{-\tau^2/M^2}d\tau +
\frac{8m\langle:\bar{\psi} \psi:\rangle}{3 M^2}\mathbf{p}^2 +
\mathcal{O}(\frac{1}{M^4}).
\end{equation}
The second term of the left hand side of the above equation is the
contribution from the more massive states, which is suppressed
exponentially and will be neglected altogether in the following.
Differentiating the above expression with respect to $\eta=M^2$, we
get the following sum rule formula for mass $m_{+}$ of spin exciton:
\begin{equation}\label{sumrule_3}
m^2_{+} = \frac{\int_{2m}^{2\Delta}\tau^2 e^{-\tau^2\eta}d\tau -
\frac{32\pi}{3}m\langle:\bar{\psi}\psi:\rangle}{\int_{2m}^{2\Delta}
e^{-\tau^2\eta}d\tau +
\frac{32\pi}{3}m\langle:\bar{\psi}\psi:\rangle\eta}.
\end{equation}

The sum rule formula for $m_{-}$ is exactly the same as $m_{+}$. In
addition, the longitudinal spin susceptibility $\chi_{zz}$ can be
analyzed similarly. When $-p^2 \gg m^2 $, the OPE result of
$\chi_{zz}(p^2,\mathbf{p})$ is
\begin{equation}\label{chi_z}
\chi_{zz}(p^2,\mathbf{p}) =
\frac{2}{\pi}\int_0^{\infty}ds\frac{\mathrm{Im}
D_0^{(0)}(s)}{s-p^2-i\epsilon}+\frac{8m_1\langle:\bar{\psi}_1
\psi_1:\rangle + 8m_2\langle:\bar{\psi}_2
\psi_2:\rangle}{3(-p^2)^2}\mathbf{p}^2.
\end{equation}
The final sum rule formula for the associated quantity $m_{z}^2$
will have the same form as Eq.(\ref{sumrule_3}) in the case of $m_1
= m_2$. Apparently, the three spin excitons are mass-degenerated.

In our trancated computation, the sum rule formula for the exciton
masses Eq.(\ref{sumrule_3}) is function of the Borel parameter $M$
or $\eta$. These two parameters are determined in the same way
presented in Sec.III. For $m_1= m_2 = 10^{-7}\Lambda$ and
$\alpha=0.8$ (SiO$_2$ case), the numerical results for the chiral
condensate are $-\langle:\bar{\psi}_1 \psi_1:\rangle =
-\langle:\bar{\psi}_2 \psi_2 :\rangle = 1.70\times
10^{-12}\Lambda^{2}$. From these quantities, the SVZ estimation for
the mass of the spin excitons is $m_{\sigma} = 0.034\mathrm{meV}$.
For suspended graphene with $\alpha = 2.2$,
$-\langle:\bar{\psi}\psi:\rangle =3.33\times 10^{-11}\Lambda^{2}$
and the estimated value for the spin exciton mass is
$0.12\mathrm{meV}$.

The typical masses obtained above are illustrated in Fig.1.
Comparing these masses, we see that the spin excitons are more
massive than Goldstone type excitons, while their masses are smaller
than $2\Delta(0)$, which is consistent with the assumption made
before Eq.(\ref{Im_chi}). It is interesting to compare our results
shown in Fig.1 with the Fig.3 of Ref. \cite{Araki2010}, where the
masses of pion-like excitons were calculated in the framework of
lattice gauge theory.

\begin{figure}\label{fig1}
\centering
\includegraphics[width=0.45\textwidth]{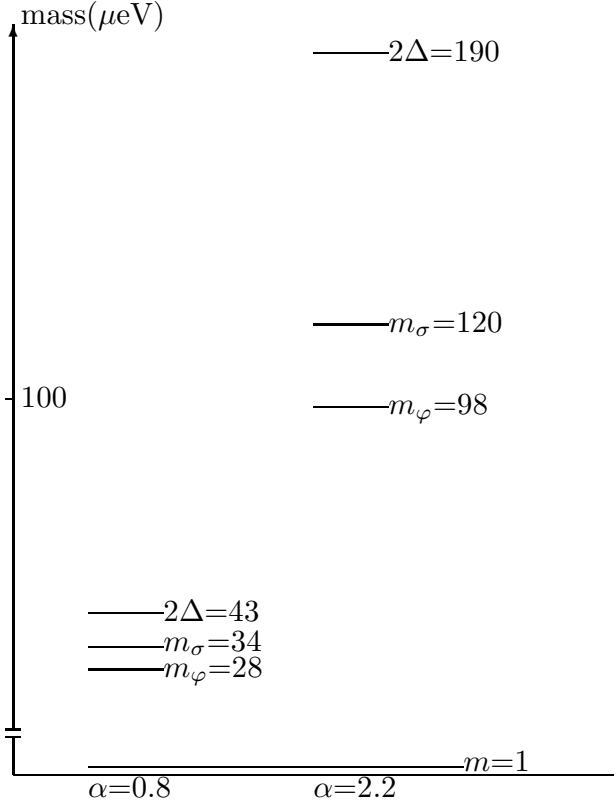}
\caption{Typical mass scales for effective fermion gap $2\Delta$,
CDW exciton $m_{\varphi}$ and spin exciton $m_{\sigma}$. The bare
fermion mass $m$ is chosen as $1 \mu\mathrm{eV}$.}
\end{figure}

\section{Summary and Conclusion}\label{Sec5}

In this paper, we studied the dynamical breaking of approximate
chiral symmetry in graphene and calculated the spectra of both
Goldstone type and non-Goldstone type excitons. In the presence of a
small bare fermion mass, the critical Coulomb interaction strength
$\alpha_c$ for dynamical fermion mass generation is reduced and the
dynamical fermion mass is substantially enhanced. In particular, we
found that the effective fermion mass ($\Delta(0) = 2.15\times
10^{-6}\Lambda$ for graphene on SiO$_2$ with $\alpha=0.8$ and
$\Delta(0) = 9.62\times 10^{-6}\Lambda$ for suspended graphene with
$\alpha=2.2$) is much larger than bare fermion mass ($m =
10^{-7}\Lambda$). Apparently, the enhancement of fermion mass
originates from the excitonic pairing instability due to Coulomb
interaction.

When the fermions have small bare mass, an approximate chiral
symmetry is dynamically broken, thus there appear massive Goldstone
excitons. In the symmetry-breaking phase, there are no massless
excitations, and all fermionic and bosonic excitations become
massive. We calculated the masses of Goldstone type excitons using
the SVZ sum rule method developed and widely used in QCD and show
that they are larger than bare fermion mass but smaller than
dynamical fermionic gap $2\Delta(0)$. In order to take into account the effect of
chiral symmetry breaking, OPE technique was used in the calculation
of two-point correlation functions. In graphene, besides Goldstone
type excitons, it is also interesting to study the non-Goldstone
type, spin excitons. We specified their positions by the same SVZ
sum rule method and found that the masses of these spin excitons are
much larger than bare fermion mass but smaller than $2\Delta(0)$.
Moreover, their masses are larger than those of Goldstone type
excitons.

In the theoretical treatment of graphene, the most widely used
approximation is to keep only the nearest hopping and expand the
fermion energy around the neutral Dirac points. Additional terms
will be included in the effective continuous field theory when
higher order corrections are taken into account. Some of these terms
may break the chiral symmetry. Generally, the symmetry-breaking
terms can appear in two classes: either as quadratic terms
$\bar{\psi}\Gamma\psi$ or as quartic terms $(\bar{\psi}\Gamma\psi)^2$
with $\Gamma=1,\gamma_3, i\gamma_5$. The former corresponds to 
fermion mass terms and through a unitary transformation of 
the 4-spinor $\psi$ the coefficients of $\bar{\psi}\Gamma\psi$ can be
absorbed into the bare fermion mass $m_s$ introduced in our Eq.(1).
Since our calculations and results depend only on the magnitude of
$m_s$, not on its physical origins, such terms will not
qualitatively affect our conclusion if they are sufficiently small.
The quartic term $(\bar{\psi}\psi)^2$ can be considered as a
Hubbard-type short-range interaction term. Its effect is more
complex than the quadratic term because we need to study the
interplay of the long-range Coulomb interaction and this short-range
interaction \cite{Liu2009}. When such short-range interaction is
weak, its contribution to the Dyson-Schwinger gap equation (3) can
be studied by the methods presented in \cite{Liu2009}. The
calculations of the masses of Goldstone bosons and non-Goldstone
bosons can be performed using the sum rule and OPE methods of
Sec.III and Sec.IV, because these calculations rely only on the
chiral condensate, which is easily obtained from the fermion gap
function. However, if this quartic interaction is strong, the whole
Lagrangian of the continuous theory is no longer chiral symmetric,
and there will not be dynamical chiral symmetry breaking and
Goldstone bosons. Indeed, our calculations are based on the
assumption that the symmetry-breaking interaction is absent or
sufficiently weak.

After obtaining the masses of various types of exciton, the next
problem is to judge whether and when these collective modes exist in
graphene. Since their masses correspond to the resonance positions
in the low energy region, we hope that NMR and neutron scattering
might be able to address this problem, similar to the efforts in
high temperature superconductors \cite{Seradjeh_Herbut2007, Ong2009,
Kee2009}. To make a connection with experiments, it is also
necessary to calculate some observable quantities that can describe
the effects of massive excitons. This issue is beyond the scope of
the present paper and will be discussed in the future.

\section{Acknowledgments}

C.X.Z is grateful to Yasufumi Araki for calling his attention to
Ref. \cite{Araki2010} and for discussions on relevant problems.
G.Z.L. thanks Juergen Dietel for helpful discussions on Kekule
distortion. C.X.Z. and M.Q.H. were supported by the National Natural
Science Foundation of China under Grant No.10975184. G.Z.L. was
supported by the National Natural Science Foundation of China under Grant
No.11074234 and the Project Sponsored by the Overseas Academic
Training Funds of University of Science and Technology of China.

\end{document}